\documentclass{aa}
\usepackage{psfig}
\usepackage{graphicx}

\def\broni{IGR J19140+0951}
\def\brone{EXO 1912+098}
\def\optical{2MASS 19140422+0952577}
\def\ecs{erg~cm$^{-2}$s$^{-1}$}
\def\lum{erg~s$^{-1}$}

\begin{document}

\title{Optical identification of \broni}

\titlerunning{Optical identification of \broni} 
\authorrunning{J.J.M. in 't Zand et al.}

\author{
J.J.M.~in~'t~Zand\inst{1,2},
P.G.~Jonker\inst{1,3},
G.~Nelemans\inst{4},
D.~Steeghs\inst{3} \&
K.~O'Brien\inst{5}
}


\institute{     SRON National Institute for Space Research, Sorbonnelaan 2,
                NL - 3584 CA Utrecht, the Netherlands 
	 \and
                Astronomical Institute, Utrecht University, P.O. Box 80000,
                NL - 3508 TA Utrecht, the Netherlands
         \and
                Harvard-Smithsonian Center for Astrophysics, 60 Garden Street,
                Cambridge, MA 02198, Massachusetts, U.S.A.
         \and
                Dept. of Astrophysics, Radboud University, P.O. Box 9010, 
                6500 GL Nijmegen, the Netherlands
         \and
                European Southern Observatory, Alonso de Cordova 3107,
                Santiago, Chile
	}

\date{Received, accepted}

\abstract{\broni\ was discovered by INTEGRAL in 2003 in the 4-100 keV
band. Observations with INTEGRAL and RXTE provide a tentative
identification as a high-mass X-ray binary (HMXB) with a neutron star
as accretor. However, an optical counterpart was thus far not
established, nor was the presence of a pulsar which is commonly
observed in HMXBs. We observed \broni\ with {\it Chandra} and find the
source to be active at a similar flux as previous measurements. The
lightcurve shows a marginally significant oscillation at 6.5~ks which
requires confirmation. We determine a sub-arcsecond position from the
{\it Chandra} data and identify the heavily reddened optical
counterpart \optical\ in the 2MASS catalog. Optical follow-up
observations with the William Herschel Telescope at La Palma exhibit a
continuum spectrum coming out of extinction above 7000~\AA\ without
strong absorption or emission features.  $V, I$ and $K_{\rm s}$ band
photometry point to an optical counterpart extincted by
$A_V=11\pm2$. The extinction is consistent with the interstellar
value. None of the data reject the suspicion that \broni\ is an HMXB
with additional circumstellar obscuration around the
accretor. }

\maketitle 

\section{Introduction}
\label{intro}

\broni\ was discovered with INTEGRAL in March 2003 (Hannikainen et
al. 2003) during observations of the nearby (1\fdg1) microquasar
GRS~1915+105. In observations that continued until May the source was
detected 70\% of the time above a threshold of 9 to 10 mCrab (20--40
keV); the brightest flux was measured during a flare peaking at
70~mCrab (Hannikainen et al. 2004).  The position is coincident with
that of \brone\ which was discovered in archival EXOSAT data by Lu et
al. (1997).  RXTE had made pointed observations of the source prior to
the report of the discovery. A public target-of-opportunity
observation was triggered in April 2002 by a detection of the source
in the {\it BeppoSAX} Wide Field Cameras (WFCs; In 't Zand et
al. 2004). RXTE observed the source again in 2003 after the INTEGRAL
detection, for 2.8 ks. Swank \& Markwardt (2003) find the source to
vary on time scales longer than 100 s and peak at 10~mCrab (2--10
keV).  $N_{\rm H}$ is fairly high at $6\times10^{22}$~cm$^{-2}$, while
the continuum can be described by a power law spectrum with an Fe-K
line of equivalent width 500 eV.

The persistent nature of the source, as suggested from the EXOSAT and
WFC detections, was confirmed with the RXTE All-Sky Monitor.  Corbet
et al. (2004) discovered a sinusoidal modulation in the flux with a
modulation depth of about 75\% and a period of 13.558$\pm$0.004~d
present since the start of observations in 1996. The persistence
supports its identification with the orbital period of \broni. This
orbital period is highly suggestive of a high-mass X-ray binary
(HMXB). An HMXB is a binary consisting of a compact object and a normal
star heavier than a few solar masses that is transferring mass to the
compact object through a wind or Roche lobe overflow resulting in
large amounts of X-ray emission.

Rodriguez et al. (2005) made a comprehensive analysis of the March-May
2003 INTEGRAL data (1.3 Msec of exposure at energies above 4 keV) and
2002-2004 RXTE observations (13 ks above 3 keV). They identify 4
states based on the observed 20--40 keV photon flux, and are able to
model the spectrum of each state with a Comptonized component of
plasma temperature 10 to 20 keV which in all but the brightest state
is complemented with a faint blackbody component.  Alternatively, the
Comptonized component can be replaced by a cutoff power law with a
photon index between 1.4 and 2.4. In almost all observations an Fe-K
line is required.  The data do not allow an accurate measurement of
the line width, although the INTEGRAL data for the middle two states
suggest a broad line of width 0.4-1.2~keV or more (1$\sigma$). The
source is detected up to $\sim100$ keV and the unabsorbed flux at
maximum is $2.1\times10^{-9}$~\ecs\ (1-20 keV) and
1.0$\times10^{-9}$~\ecs\ (20-200 keV). The RXTE data show a variable
$N_{\rm H}$ between 3 and 10$\times10^{22}$~cm$^{-2}$, bringing
\broni\ marginally in the range of the 'obscured INTEGRAL sources'
(for a recent review, see Kuulkers 2005). The Galactic $N_{\rm H}$
measured by Dickey \& Lockman (1990) within 1\degr\ from the source is
between 1.3 and $1.9\times10^{22}$~cm$^{-2}$; the data point closest
to the source (at 0\fdg24) is $1.8\times10^{22}$~cm$^{-2}$.  Rodriguez
et al. find that the spectroscopic evidence favors a neutron star over
a black hole for the nature of the accretor: in the 20--200 keV
luminosity versus 1--20 keV luminosity diagram (first presented by
Barret et al. 1996), the source is in the X-ray burster domain as long
as the distance is below 10~kpc.

Thus, all evidence points to \broni\ being an HMXB with strong and
variable local absorption and a neutron star as accretor. Prior to the
INTEGRAL discovery we had already proposed a {\it Chandra} observation
of \brone, based on the WFC detection, to search for possible long
X-ray pulsations in a faint source and determine an accurate position
to search for the optical counterpart. The data were obtained in
2004. They now provide an opportunity to make a conclusive
identification of \broni\ as an HMXB.  Here we present the results of
this observation and of the optical follow up.

\begin{figure}[!t]
\centering
\includegraphics[width=0.9\columnwidth]{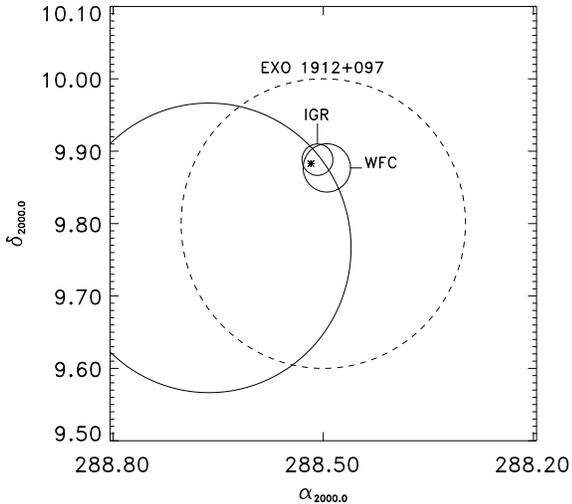}
\caption{Map showing source detections around \broni\ \label{k1914}.
There are two positions for EXO 1912+097. The dashed circle refers to
the coordinates as provided by Simbad, the solid circle to those
inferred from Lu et al. (1997). The INTEGRAL data (JEM-X and IBIS
combined) are from Cabanac et al. (2004). The asterisk designates the
{\it Chandra} position. The WFC error circle is from In 't Zand et al. (2004).
Circle radii are for a 90\% confidence level.}
\end{figure}

\begin{figure}[!t]
\centering
\includegraphics[height=0.9\columnwidth,angle=270]{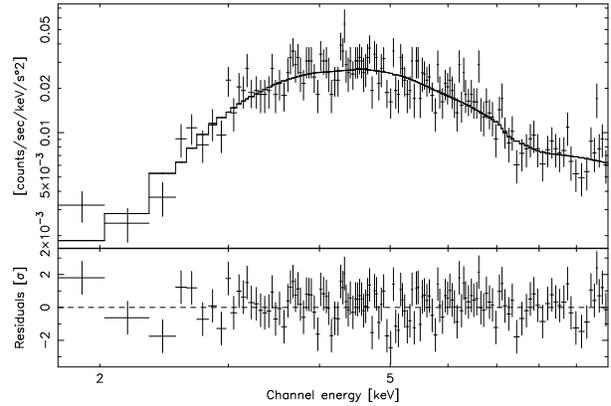}
\caption{{\it (Top)} {\it Chandra} spectrum (crosses) as derived from
the PSF with the best fit power law model (histogram). {\it (Bottom)}
residuals with respect to the best fit model.\label{chsp}}
\end{figure}

\section{{\it Chandra} observation}

{\it Chandra} observed \broni\ on May 11th, 2004, starting at 17:53:30
UT, with the ACIS-S CCD array (Garmire et al. 2003) in the focal plane
and no grating. The CCD frame time of the data is 3.2~s, the exposure
time 20.10 ks. The binary orbital phase coverage is 0.833--0.850 with
an uncertainty of 0.055 based on the ephemeris of Corbet et
al. (2004). Thus, the observation was well ahead of the maximum
orbital modulation. The source is clearly detected. It is the only
source brighter than 0.0003 c~s$^{-1}$ inside the INTEGRAL error
circle (Hannikainen et al.  2004). The image shows a piled-up source
with readout trails but no obvious hole at the center of the point
spread function. The average position of all photons within 10\arcsec\
is $\alpha_{2000.0}=19^{\rm h}14^{\rm m}4\fs232$,
$\delta_{2000.0}=+9^\circ 52\arcmin58\farcs29$ ($l^{II}=44\fdg30$,
$b^{II}=-0\fdg47$) with a nominal uncertainty of 0\farcs6.  This
position is consistent with that of \brone, \broni\ and the {\it
BeppoSAX}/WFC position, see Fig.~\ref{k1914}.

\begin{figure}[!t]
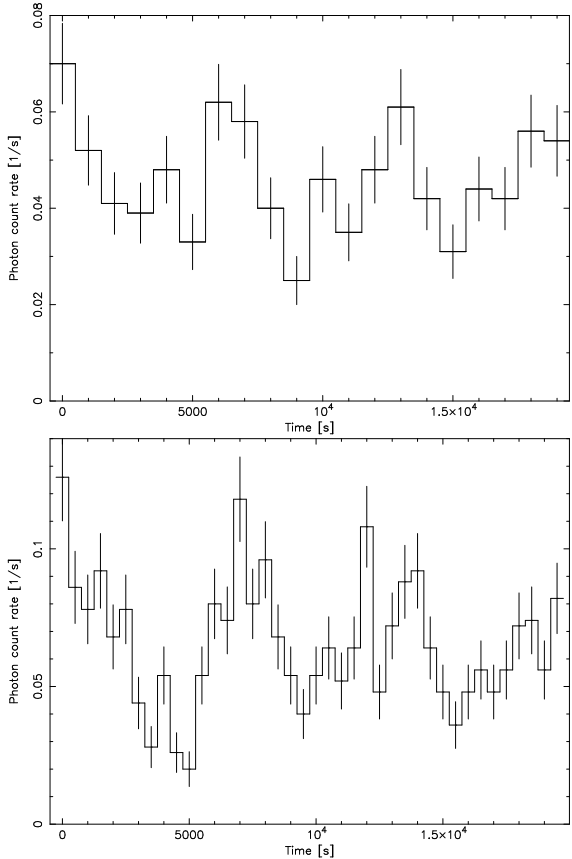

\centering
\includegraphics[height=0.85\columnwidth,angle=270]{3411f3a.ps}
\includegraphics[height=0.85\columnwidth,angle=270]{3411f3b.ps}
\caption{Light curves of \broni\ during the {\it Chandra} observation
in full bandpass.  {\it (Top)} From photons in the readout trail
(resolution 1000 s) {\it (Bottom)} From photons within an annulus with
an inner radius 2\arcsec and outer radius 5\arcsec\ (resolution 500
s).
\label{chandra_lc}}
\end{figure}

We reprocessed the level-1 data with CIAO version 3.2.1 to resurrect
afterglow events, using among other tools {\tt
acis\_run\_hotpix}. This procedure adds 6.9\% to the number of photons
within 2\farcs0 of the source. We extracted a source spectrum from the
photons in this region with the CIAO tool {\tt psextract}; a
background spectrum was extracted from a circular region with a radius
of 1\arcmin\ on the same CCD and centered 70\arcsec\ from the source
centroid. The spectral analysis was performed with {\tt XSPEC} version
11.3.1 (Arnaud 1996). We employed the {\tt pileup} model in {\tt
XSPEC} which was specifically designed for the analysis of ACIS data
(Davis 2001), leaving free the morphing parameter $\alpha$ and the
fraction of the point-spread function (PSF) that is piled up. The
spectrum was binned so that at least 15 photons are contained per bin,
thus allowing the use of the $\chi^2$ statistic as a goodness of fit
estimator. The resulting spectrum can be well fitted with an absorbed
power-law ($\chi^2_\nu=1.005$ for 130 degrees of freedom). The photon
index is $\Gamma=1.1\pm0.8$ (all errors quoted in this paper are for a
90\% confidence level) and $N_{\rm
H}=(1.0\pm0.3)\times10^{23}$~cm$^{-2}$. These values are similar to
those found by Rodriguez et al. (2005). The absorption column is
considerably larger than the interstellar value. Figure~\ref{chsp}
shows the spectrum and model fit. Very few photons are detected below
1 keV and they were not taken into account when fitting models,
neither were photons above 10 keV. The unabsorbed 1-10 keV flux is
$3.9_{-1.5}^{+3.0}\times10^{-11}$~\ecs, but this number is susceptible
to systematic errors because of the uncertainty of the pileup
correction. Therefore, we verified it with the spectrum of the readout
trail. This spectrum was extracted with the CIAO tool {\tt
acisreadcorr}. A background spectrum was extracted from a strip on the
detector 50\arcsec\ to the north.  The spectral shape is consistent
with the above result: $\Gamma=0.7\pm0.7$, $N_{\rm
H}=(0.8\pm0.3)\times10^{23}$~cm$^{-2}$ ($\chi^2_\nu=0.96$ for 51
degrees of freedom). If we fix the spectral shape to the more
accurately determined former shape and fit the flux, we find
$(1.5\pm0.2)\times10^{-10}$ (1-10 keV) or
$(4.0\pm1.0)\times10^{-10}$~\ecs\ (1-20 keV). This is fainter than in
all INTEGRAL detections analyzed by Rodriguez et al. (2005) but 50\%
brighter than in the brightest pointed RXTE observation.  We find no
evidence for an Fe-K line at 6.4 keV.  Adding a Gaussian line at
6.4~keV improves the fit by $\Delta\chi^2=4.5$ in the PSF spectrum
which is not significant (10\% chance probability).  The
90\%-confidence upper flux limit on a narrow line is
$7\times10^{-5}$~phot~s$^{-1}$cm$^{-2}$; on a Gaussian line with
$\sigma=0.4$~keV it is $3.7\times10^{-4}$~phot~s$^{-1}$cm$^{-2}$. The
limits on the equivalent widths are 116 eV and 839 eV
respectively. These values are consistent with the INTEGRAL and RXTE
detections for a broad line, but inconsistent for a narrow line. Thus,
a comparison of the {\it Chandra} with the INTEGRAL/RXTE observations
rules out a narrow line, if the line is present with a similar
equivalent width.

We generated light curves from the trailed photons at various time
resolutions between 40 $\mu s$ and 1000~s. The high-resolution data do
not show interesting features. We show in Fig.~\ref{chandra_lc} (top)
the time profile at 1000~s resolution.  The source is obviously
variable. There is even a hint of an oscillation with a period of
about 6.5 ks. We also generated a light curve from photons in an
annulus around the PSF (Fig.~\ref{chandra_lc} bottom), thus minimizing
pile up effects. The modulation is also visible here.  We checked RXTE
All-Sky Monitor (ASM) data for a confirmation. The ASM data are
marginally qualified for that because the detection threshold is about
40\% , and the time resolution is high enough (90~s) to resolve a
6.5~ks modulation. We tested the full data set as well as orbital
phases at which the flux is highest and the three energy channels
separately. The times were corrected for the Earth's orbit around the
Sun, but not for the satellite or binary orbit which are not important
for a possible 6.5~ks pulsation. We are unable to obtain an
unambiguous confirmation from the ASM data. This negative ASM result
implies that the 6.5~ks modulation probably is not related to a
neutron star spin.  However, additional confirmation may be worthwhile
since modulation patterns may change (e.g., Patel et al. 2004).

\section{Optical and near-infrared observations}

\begin{figure*}[!t]
\centering
\includegraphics[width=\columnwidth]{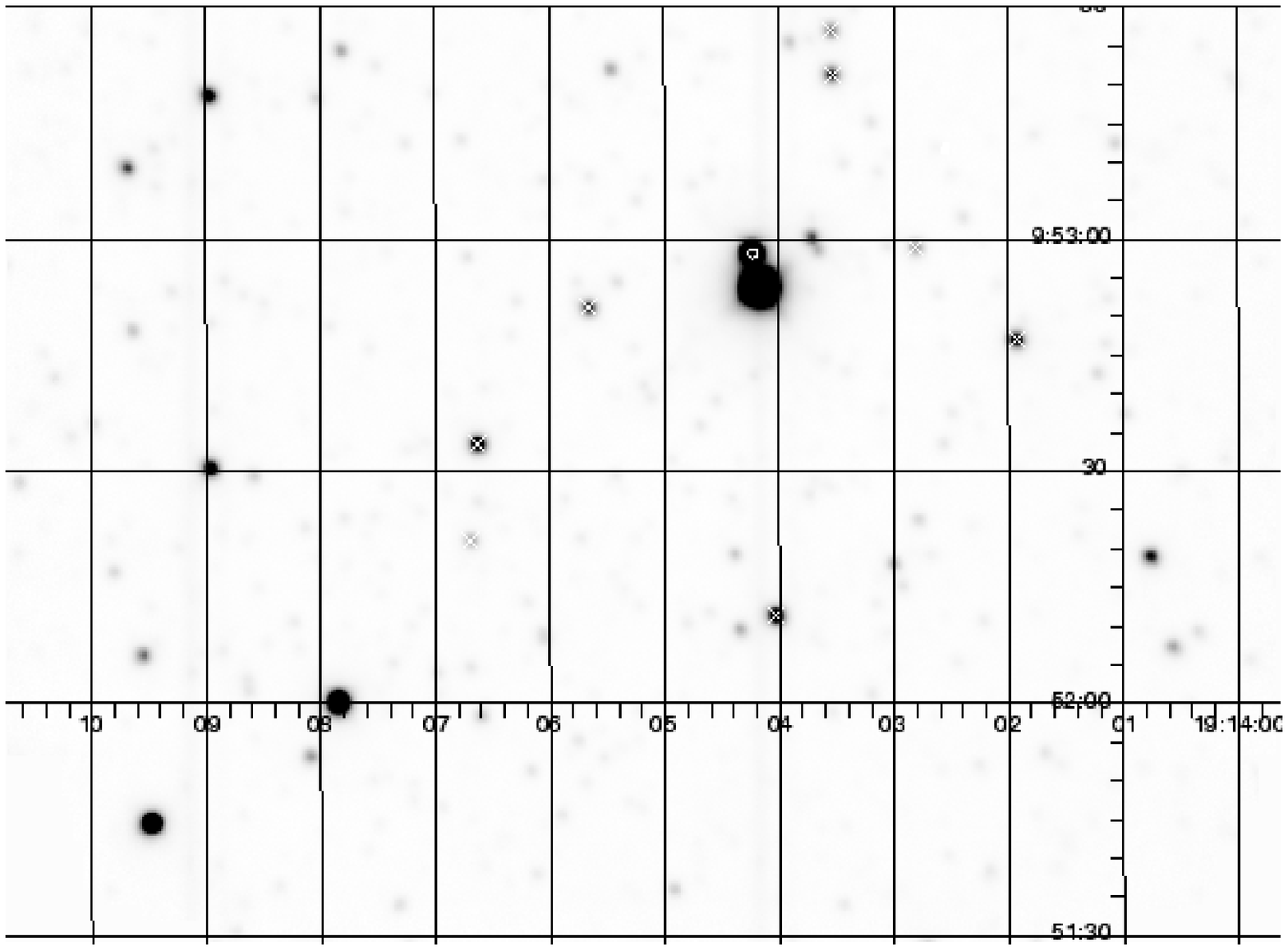}
\includegraphics[width=\columnwidth]{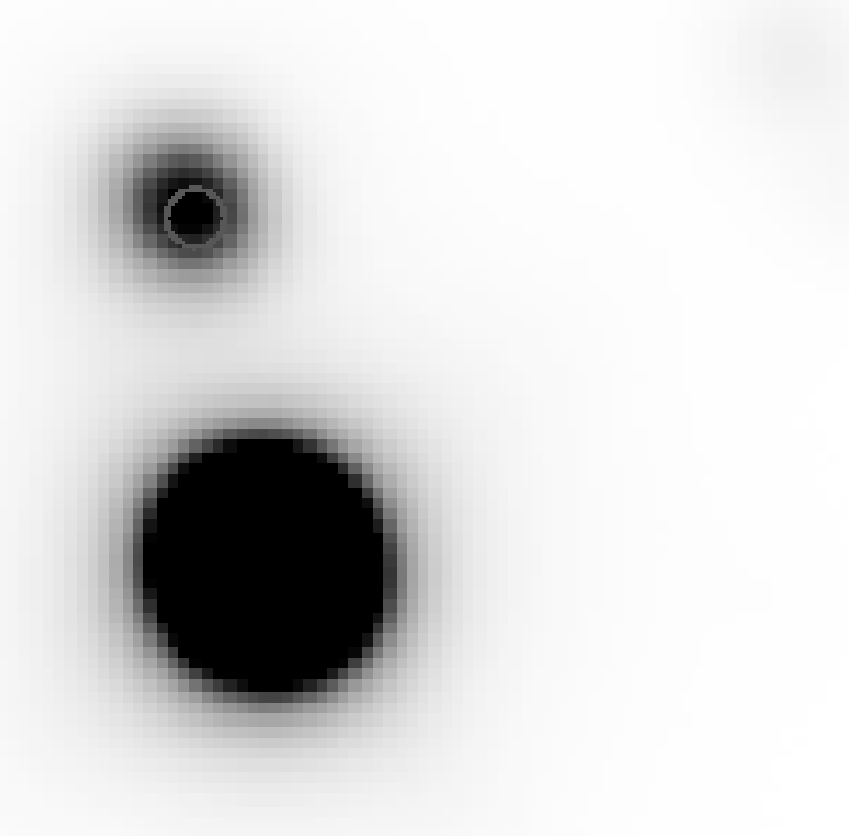}
\caption{{\it (Left)\/} $K_{\rm s}$-band image with the
0\farcs6-radius {\it Chandra} error circle. The stars designated with
white crosses were used for the astrometric solution through cross
correlation with the 2MASS catalog. This image was taken with SOFI on
the NTT at the ESO on May 18, 2005, under non-photometric
conditions. The seeing was 1\farcs3 (FWHM). {\it (Right)\/}
15\arcsec$\times$15\arcsec zoomed in image with an adapted grey scale.
\label{2mass}}
\end{figure*}

\subsection{Identification of counterpart}

We cross checked the {\it Chandra} position with the USNO B1.0 (Monet
et al.  2003) and 2MASS (Skrutskie et al. 1997) catalogs and found a
2MASS object consistent with the {\it Chandra} position (\optical). It
is fairly close (4\farcs0) to another bright object to the south. We
acquired a near-infrared image of the field in the $K_{\rm s}$ band
under non-photometric conditions using the 3.6~m ESO NTT with SOFI,
see Fig.~\ref{2mass}. It clearly reveals the two objects near the {\it
Chandra} position. After calibrating the plate scale through 9
isolated stars that were identified in the 2MASS catalog, the northern
object was localized at $\alpha_{2000.0}=19^{\rm h}14^{\rm m}4\fs231$,
$\delta_{2000.0}=+9^\circ 52\arcmin58\farcs35$ which is 0\farcs06 from
the Chandra position of the X-ray source and 0\farcs6 from the 2MASS
position of the optical counterpart. The southern object is present in
the USNO B1.0 catalog while the northern is not, testifying to the
redness of the optical counterpart. The 2MASS magnitudes are $K_{\rm
s}=7.06\pm0.20$ for the northern and $6.274\pm0.026$ for the southern
object.

\subsection{Spectroscopy}

\begin{figure}[!t]
\centering
\includegraphics[height=\columnwidth,angle=270]{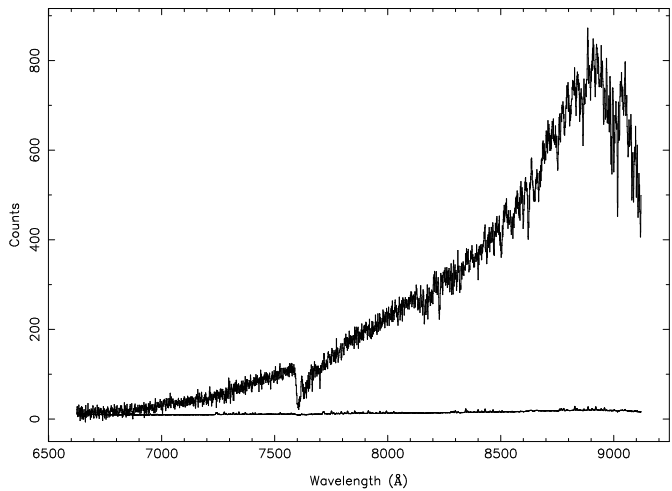}
\caption{Optical spectrum of \optical, taken with the red arm of the
ISIS spectrograph on the WHT at La Palma. The background was
subtracted, but no flux calibration was performed because no
observations of standard stars were performed. The instrument response
is fairly flat between 5500 and 8500~\AA\ after which it drops by 60\%
at 8900~\AA\ and by 98\% at 9500~\AA. Thus, this spectrum illustrates
the strong reddening.
\label{opticalspectrum}}
\end{figure}

We obtained two spectra of \optical\ on August 12, 2004, each with an
exposure time of 1500~seconds, using the ISIS spectrograph mounted on
the 4.2~m William Herschel Telescope located at the Roque de Los
Muchachos Observatory, La Palma, Spain. We used the R316R grating with
the central wavelength set to 7500~\AA\ and a slit width of 1\arcsec\
in combination with the MARCONI2 CCD. This setup yields a mean
dispersion of 0.83~\AA\ per pixel.  The source was observed at an
airmass of $\sim$1.1. It, as well as the southern source in the 2MASS
image, is not resolved into multiple sources with a seeing of 1\farcs2
(FWHM). Due to the high interstellar extinction towards the source we
are not able to detect the source spectrum in the blue arm of the ISIS
spectrograph. In the red arm of the spectrograph we detect the source
redward of $\approx 7000$~\AA, see Fig.~\ref{opticalspectrum}. We find
weak evidence for absorption lines from the Paschen series. However,
due to the relatively low signal--to--noise ratio, the presence of
telluric lines at or close to the wavelength of some of the Paschen
lines, and fringing in the red part of the CCD for which no
calibration measurements are available, the continuum level is too
uncertain to determine. We estimate a conservative upper limit to the
equivalent width of any narrow line of 1.6~\AA.  The spectroscopic
data have too narrow a bandpass to confidently determine the spectral
type. The presence of Paschen absorption lines, if confirmed, would
point to an A, B or late O--type companion star.

\subsection{Photometry}

The 2MASS magnitudes for the optical counterpart are $J=8.55\pm0.05$,
$H=7.67\pm0.16$ and $K_{\rm s}=7.06\pm0.20$.  However, this is based
on aperture photometry without taking into account the blending by the
southern source. Therefore, the brightnesses are overestimated. This
is shown through a measurement of $K_{\rm s}$ using the new image
taken on May 18, 2005, albeit under non-photometric circumstances
(Fig.~\ref{2mass}). After calibrating the magnitude scale with 4
isolated stars that are also present in the 2MASS catalog and have the
same magnitude differences among them, we find $K_{\rm s}=8.69\pm0.05$
which is 1.6 mag fainter than in the 2MASS catalog.

We obtained additional images in the $V$ (180 s total exposure) and
$I$ band (240~s) with the 1~m~Henrietta Swope telescope at Las
Campanas Observatory on August 7, 2005. The seeing was 1\farcs3 at an
airmass of $\sim$1.3. The counterpart was detected in $I$ at
13.0$\pm0.1$ mag, while it was not detected in $V$. A limit of
$V>18.8$ mag was determined from the magnitude of the faintest
unambiguous star visible in the field.

We also checked infrared data taken with the Midcourse Space
Experiment (MSX; Mill et al. 1994) and find a $0.281\pm0.013$~Jy
object in the most sensitive 8.3~$\mu$m band at
$\alpha_{2000.0}=19^{\rm h}14^{\rm m}4\fs16$,
$\delta_{2000.0}=+9^\circ 52\arcmin55\farcs4$ with a 1-sigma
uncertainty of 0\farcs8. This is 2\farcs5 from the optical counterpart
(or 3$\sigma$) and 1\farcs6 (2$\sigma$) from the nearby southern
object. Given the uncertainty we cannot rule out an association with
the unrelated southern object.

\begin{figure}[!t]
\centering
\includegraphics[width=\columnwidth]{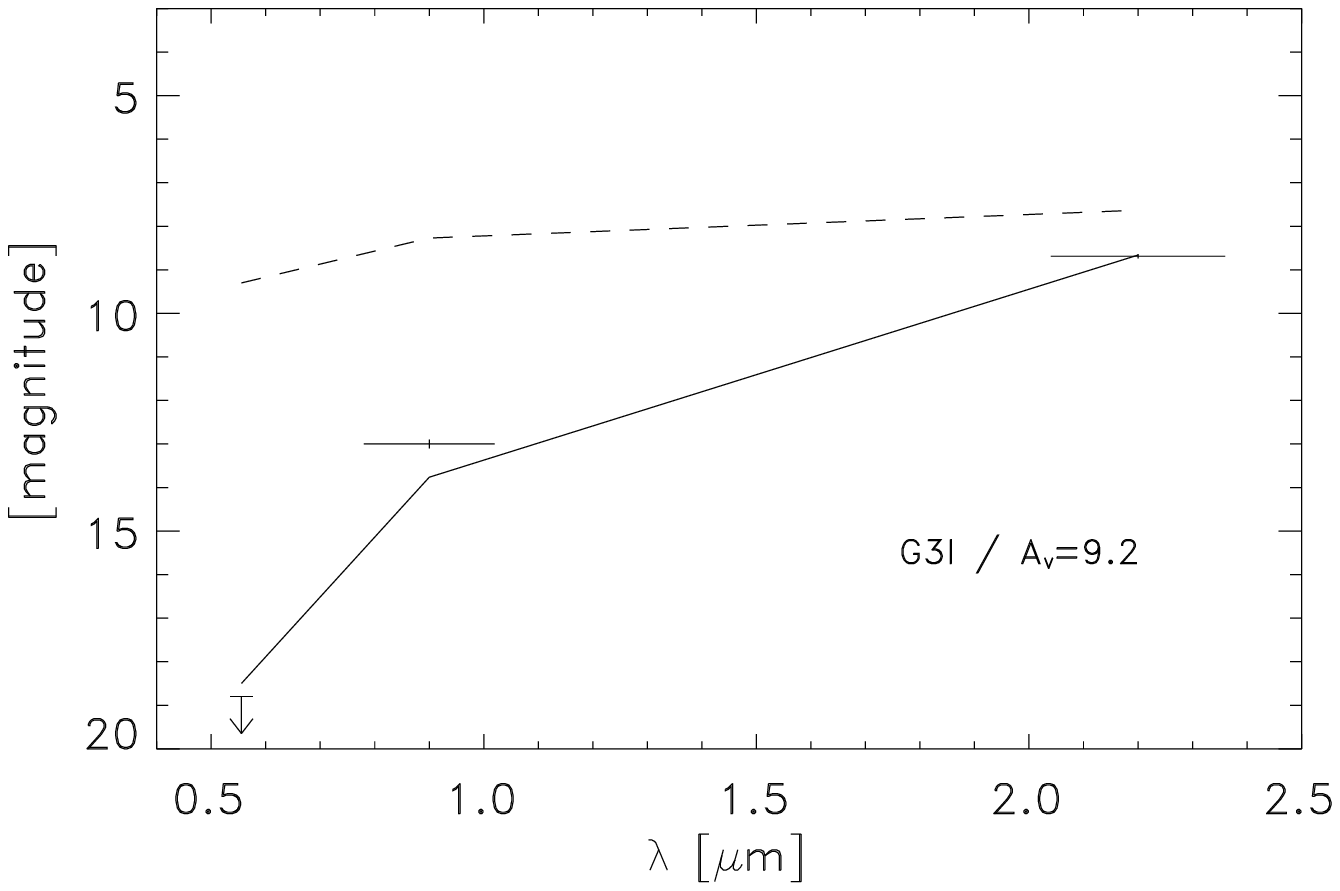}
\includegraphics[width=\columnwidth]{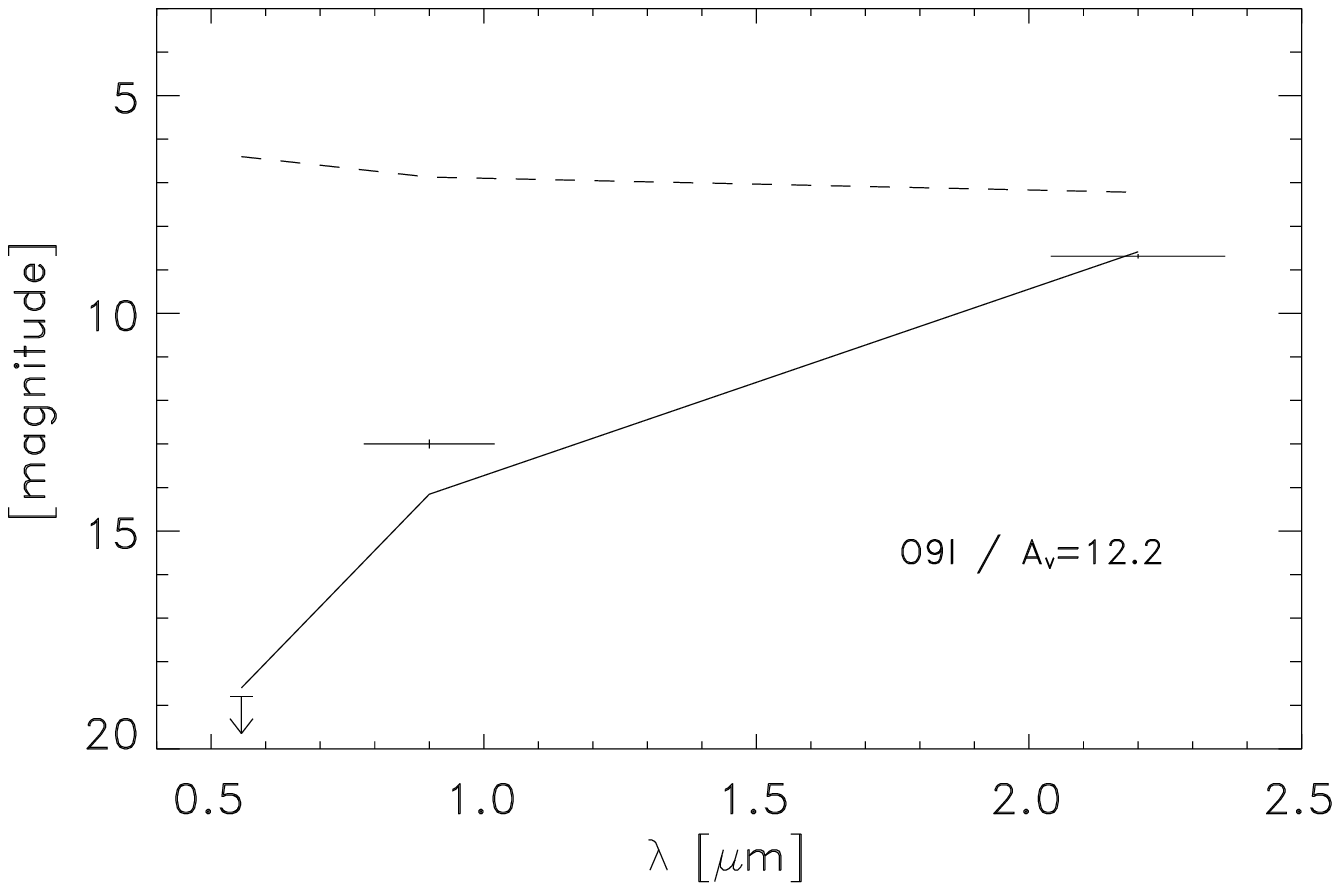}
\caption{Photometry in $V, I$ and $K_{\rm s}$ (crosses; horizontal
bars indicate band width and vertical bars magnitude errors), fitted
model (solid line) and fitted model before extinction (dashed line).
The top panel shows the best fit with spectral type G3I and $A_V=9.2$;
the bottom panel shows, as example, the fit with a O9I model leaving
free only $A_V$. Both models are unacceptable, but $\chi^2_\nu$ is two
times larger for the O9I model. The models were taken from Cox (2000)
and the extinction laws from Schlegel et al. (1994).
\label{figphotfit}}
\end{figure}

We sought confirmation on the suspected early spectral type by fitting
photometric models for main sequence and supergiant stars (Cox 2000),
using extinction laws from Schlegel et al. (1994), to the $VIK_{\rm
s}$ photometry. The results are ambiguous, see
Fig.~\ref{figphotfit}. The best fit models are late-type stars (G3I or
V), but the fits are not acceptable. Photometric fits with earlier
spectral types are worse.  Possibly the non-simultaneity of the
$K_{\rm s}$ measurement with the $V$ and $I$ measurements plays a
role.

The extinction leaves an unambiguous imprint on the photometry, and is
fairly independent of spectral type. The values for $A_V$ range
between 9 and 12 and are, based on the translations determined by
Predehl \& Schmitt (1995), consistent with $N_{\rm
H}=(1.9\pm0.3)\times10^{22}$~cm$^{-2}$.

\section{Discussion}

Our observations result in the secure identification of the optical
counterpart to one of the 'obscured INTEGRAL sources' (e.g., Kuulkers
2005). Like the counterpart of the prototypical IGR J16318-4848
(Filliatre \& Chaty 2004), it is a peculiar highly reddened object
with bright NIR magnitudes whose spectral type still needs to be
resolved.

The most likely scenario is that the optical counterpart is an
early-type star in an HMXB. In analogy to other HMXBs, the X-ray
persistence, the orbital modulation and the lack of a short pulse
period suggests a supergiant donor star. An alternative to an HMXB
scenario would be a low-mass X-ray binary (LMXB) where a Roche-lobe
filling sub-solar mass star is donating matter to a compact object via
an accretion disk. This is unlikely on two grounds. First, the orbital
period is much longer than most LMXBs (c.f., Liu et al. 2001).
Second, optical counterparts to LMXBs typically have $M_{K_{\rm s}}$
between -2 and 2 (e.g., Wachter et al. 2005). This would imply for
\broni\ a distance between 30 and 90 pc and a maximum 1-200 keV
luminosity of $3\times10^{33}$~\lum. This is inconsistent with its
persistently active behavior: accretion disks become unstable at such
low luminosities for the orbital period measured for \broni\ (e.g.,
Van Paradijs 1996).

Support for the HMXB scenario also comes from the position: \broni\ is
in the direction of the tangent to the Sagittarius arm.  HMXBs are
expected to be coincident with spiral arms because that is where young
systems reside. If the association with the Sagittarius arm is true,
the distance is of the order of 2 to 6 kpc. The implied 1--20 keV
luminosity during the {\it Chandra} observation is roughly
10$^{35}$~\lum, a common value for an HMXB.

There is no evidence for obscuration of the optical counterpart beyond
the interstellar values provided by Dickey \& Lockman (1990), while
there is such evidence for the accretor through X-ray
observations. The orbital modulation of the X-ray flux (Corbet et
al. 2004) suggests either the presence of an inclined circumstellar
disk formed by the donor or an eccentric orbit, or both.

Further understanding of \broni\ would benefit from a longer {\it
Chandra} or {\it XMM-Newton} observation, to study the variability (in
particular possible pulsations) without data gaps due to earth
occultations, and from a calibrated infrared spectrum in the I to L
bands, in order to measure spectral lines and achieve a reliable
spectral classification.

\acknowledgement 

We thank Mariano M\'{e}ndez for useful discussions and the anonymous
referee for useful suggestions. JZ acknowledges support from the
Netherlands Organization for Scientific Research (NWO).  DS
acknowledges a Smithsonian Astrophysical Observatory Clay fellowship.
We thank the Observatories of the Carnegie Institution of Washington
for providing access to the Swope telescope at Las Campanas
Observatory in Chile. This research has made use of the NASA/IPAC
Infrared Science Archive, which is operated by the Jet Propulsion
Laboratory, California Institute of Technology, under contract with
NASA, and of the RXTE/ASM archive provided by ASM teams at MIT and at
the RXTE SOF and GOF at NASA/GSFC.

\end{document}